\documentclass{iopart}

\usepackage{latexsym}
\usepackage{hyperref}
\usepackage{longtable}
\usepackage{array}
\usepackage{dcolumn}
\usepackage{epsfig}
\usepackage{amsfonts}
\usepackage{bm}
\usepackage{graphicx}

\newcommand{\be}{{\bm{e}}}
\newcommand{\beps}{{\bm{\eps}}}

\newcommand{\bX}{{\bm{X}}}

\newcommand{\bnull}{{\bm{0}}}

\newcommand{\bk}{{\bm{k}}}

\newcommand{\bQ}{{\bm{Q}}}

\newcommand{\al}{\alpha}

\newcommand{\eps}{\epsilon}

\begin{document}

\title{Finite strain Landau theory of high pressure phase transformations}

\author{W.~Schranz$^1$, A.~Tr\"oster$^1$, J.~Koppensteiner$^1$ and 
R.~Miletich$^2$}
\address{$^1$Faculty of Physics, University of Vienna,
Boltzmanngasse 5, A-1090 Wien, Austria}
\address{$^2$ Mineralogisches Institut, Universit\"at Heidelberg, 
Im Neuenheimer Feld 236, 69120 Heidelberg, Germany}

\ead{wilfried.schranz@univie.ac.at}

\begin{abstract}
The properties of materials near structural phase transitions are often 
successfully described in the framework of Landau theory. While usually the 
focus
is on phase transitions, which are induced by temperature changes approaching 
a critical 
temperature T$_c$, here we will discuss structural 
phase transformations driven by high hydrostatic pressure, 
as they are of major importance for understanding processes in the interior 
of our earth. 
Since at very high pressures the deformations of a material are generally very 
large, one needs to apply a fully nonlinear description taking physical as 
well as geometrical nonlinearities (finite strains) into account.  
In particular it is necessary to retune conventional Landau theory 
to describe such phase transitions.
In [A.~Tr\"oster, W.~Schranz and R.~Miletich, Phys.~Rev.~Lett.{\bf 88}, 55503
(2002)]
we have constructed a Landau-type free energy based on an 
order parameter part, 
an order parameter-(finite)strain coupling and a nonlinear elastic term. 
This model provides an excellent and efficient framework for the 
systematic study of phase transformations for a wide range of materials up to 
ultrahigh pressures.\\
We illustrate the model on the specific example of $BaCr(Si_4O_{10})$, showing 
that it fully accounts for the elastic softening which is observed near the 
pressure induced phase transformation.   
\end{abstract}

\maketitle 

\section{Introduction}

A major achievement of Ekhard Salje and coworkers was to show that Landau
theory can be very efficiently applied for the desription of experimental data
near temperature induced structural phase transitions in minerals 
(Salje, 1990; Salje, 1992; Carpenter, Salje and Graeme-Barber, 1998). 
There are a couple of reasons for this success.
First of all, as coupling to strains induce long range forces and enhance 
anisotropy,
strain-induced interactions always tend to suppress fluctuations,
leading to a mean field type of transition (Bratkovsky, 1995, Tsatskis, 1994). 
Depending on the nature of the 
system (i.e.~on the type of coupling between strain and order parameter
its anisotropy and the boundary conditions imposed on its surface) 
strain effects also may change the character of the 
transition from second to first order (Bergman and Halperin, 1976). 
Now in principle, a Landau potential can be obtained from a low order Taylor
expansion of the mean field free energy. The problem of how to determine
the range of validity of this expansion naturally arises.
To answer this question one observes that in many mineral systems
the phase transitions turn out to be of the displacive type (Dove, 1997).
For such systems, the above authors were able to show that
Landau theory indeed allows to reproduce experimental data in an 
extremely broad temperature range, in particular down to very low temperatures
by taking into account the effect of quantum saturation 
(Salje, Wruck and Thomas, 1991).
The situation is not so simple for the other extreme, i.e.~order-disorder 
phase transitions, since it is 
no longer possible to approximate their free energy 
by a low order polynomial in the order parameter  
(Giddy, Dove and Heine, 1989). In passing, we note that
real systems are often of mixed, i.e.~displacive/order-disorder character
(Meyer, 2000; Sondergeld, 2000), 
and the characterization of crossover systems 
in between a displacive and order-disorder type is still 
a matter of active research 
(Perez-Mato, 2000; Rubtsov, Hlinka and Janssen, 2000; Tr\"oster, Dellago and 
Schranz, 2005).\\
In ``traditional'' Landau theory it is difficult to include pressure effects 
beyond the infinitesimal strain approximation. Therefore one frequently resorts to 
replacing the elastic background energy by its harmonic truncation and 
consequently treats the  strains as infinitesimal. Since the total strains 
appearing in temperature induced phase transitions are as a rule quite small
this is well justified. However, when we discuss the role of extremely high pressure 
in driving the transition, the validity of this assumption must be seriously doubted.    

At high pressure a phase transition is usually detected by anomalies 
e.~g.~in the system volume $V(P)$, which in turn result from
anomalies in the pressure dependence of the lattice parameters
$a_i(P)$, $i=1,2,3$, near a critical pressure $P_c$. In many 
cases this leads to an additional (with respect to the "background") nonlinear 
$P$-dependence of lattice parameters and/or volume. 
As a function of pressure, the "stiffness" of any solid is characterized by the isothermal compressibility 
$\kappa(P)=-d \log V(P)/d P$. Various theoretical concepts
are usually employed to derive so-called equations of state (EOS)
(Anderson, 1995; Angel, 2000a; Angel, 2000b),
which -- in absence of phase transitions -- describe the hydrostatic 
pressure dependence of the crystal's volume $V(P)$.

In presence of a high pressure phase transition (HPPT) a commonly adopted practice
is to merely fit the corresponding $V(P)$-behavior 
to a number of differently parametrized EsOS 
for each phase (Schulte and Holzapfel, 1995; Kr\"uger and Holzapfel, 1992; 
Chesnut and Vohra, 2000), and, apart from computer simulations,
in the case of a reconstructive phase transition
there is little more one can do at present. However, from a theoretical point of view,
although this purely phenomenological procedure does fit experimental volume data in many
cases, it neither describes the pressure behaviour of
individual strain components nor provides any possible further insight
into the mechanisms of a HPPT. A more profound theoretical approach is therefore  of vital interest to a
broad audience
reaching form physicists studying the high pressure behavior of materials 
(crystals, liquid
crystals, complex liquids, biological membranes, etc.) to geologists
investigating the earth's minerals and bulk properties.

Indeed, for HPPT's of the group-subgroup type, it is natural to expect that one could do much better
by applying the concepts of Landau theory. In particular, we should be able to connect 
the high and low pressure phase by a \emph{uniform} thermodynamic description.
However, here one faces a marked difference compared to the temperature driven case.
At high pressures, as the inter-atomic forces opposing further
compression, a crystal's volume and lattice parameters develop
large strains in a pronounced non-linear way, which corrupts any serious attempt to 
treat the elastic energy in the framework of the infinitesimal strain approximation. 
For these reasons it is necessary to reconsider the construction of Landau theory at extremely 
high pressures and large deformations (Tr\"oster, Schranz and Miletich, 2002).
Here we summarize the main ideas of this new theoretical approach and discuss
some consequences for the elastic behaviour of materials.

\section{New approach for high pressure phase transformations}

\begin{figure}[h] 
\includegraphics[scale=0.5,angle=-90]{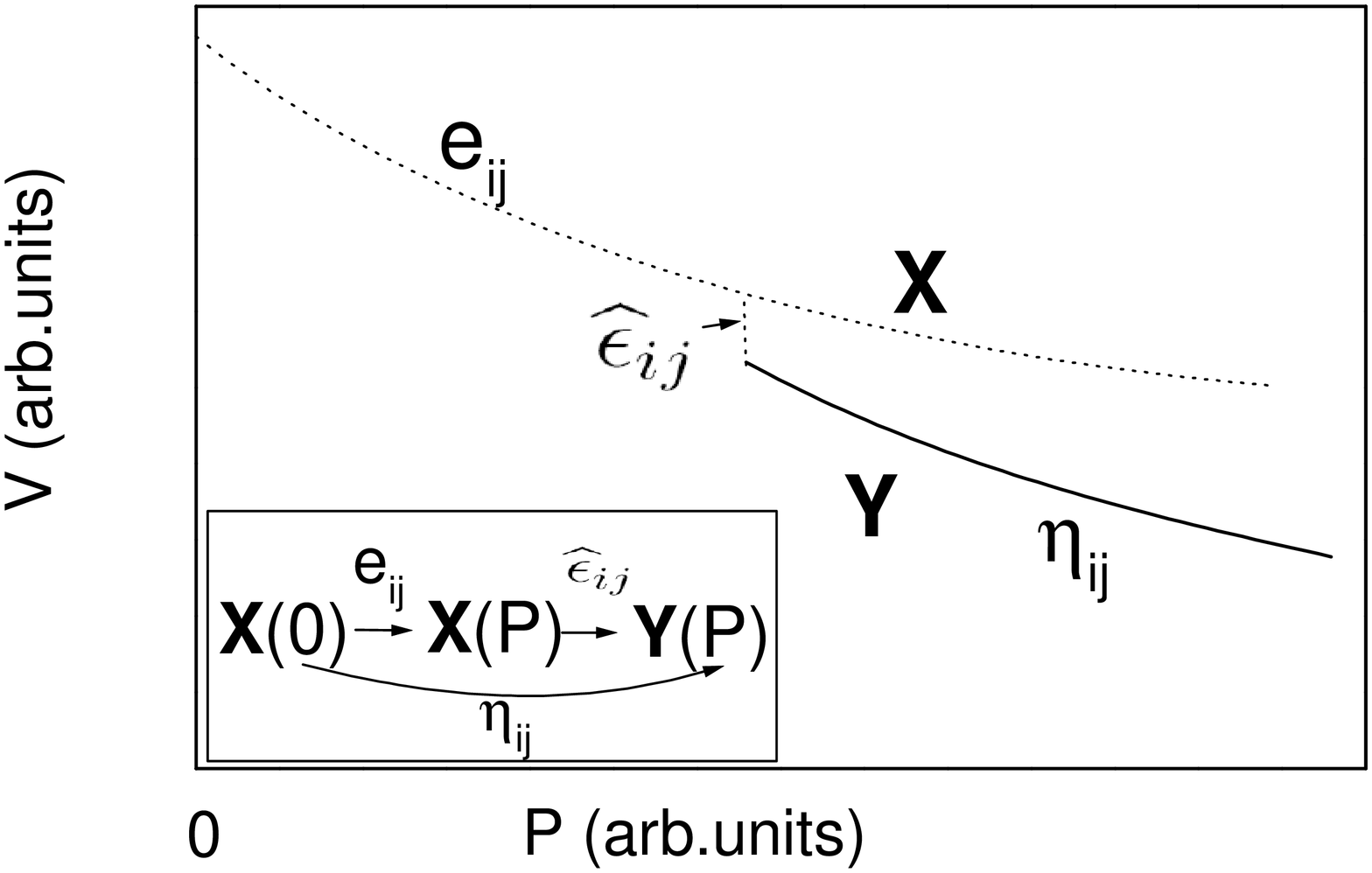}
\caption{Sketch of deformation states and superposition of strains.}
\label{Fig1}
\end{figure}

As stated above, the conventional infinitesimal strain approach is clearly insufficient
for aiming at an accurate description of HPPT's. Instead, at high pressures
the elastic response of a system must be characterized in terms of a appropriate nonlinear strain measure
like the Lagrangian or Eulerian strain tensor.
Let $\bX$ and $\bX'$ denote an undeformed reference system and a deformed system, respectively.
Then, from the deformation gradient tensor  $\alpha_{ik}=\partial X_i'/\partial X_k$
one constructs the Lagrangian strain tensor 
$e_{ik}=\frac{1}{2}\left(\sum_n \alpha_{ni}\alpha_{nk}-\delta_{ik}\right)$.
In the present work, we will consider three pairs of systems $\bX,\,\bX'$
at hydrostatic external pressures $P$ (Fig.\ref{Fig1}). For measuring the 
``background strain'' $\be(P)$, which is defined as the Lagrangian strain displayed
by the system at the constraint of zero order parameter $\bar\bQ$ (see below), $\bX(P)$ denotes the
corresponding state and $\bX(P=0)$ is used as a reference system. 
For measuring the spontaneous strain $\bm{\widehat\eps}$ resulting from
the emergence of $\bar\bQ\ne\bnull$, $\bX(P)$ plays the role of a (hypothetical) ``floating''
background reference system. On the other hand, above the critical pressure $P_c$ the total strain
$\bm{\eta}(P)$, which is obtained
as the nonlinear superposition $\bm{\eta}=\be+\bm{\al}^+.\widehat{\bm{\eps}}.\bm{\al}$. 
of the spontaneous strain ${\widehat\beps}$ appearing ``on top of'' 
the background strain $\be(P)$, is the strain measured in the fully deformed state 
$\bm{Y}(P)$, which is the one actually displayed by the system, the system $\bX(0)$ again
playing the role of the reference system.

In the spirit of Landau's theory of phase transitions the Landau free energy
$F({\bQ, \beps})$ is built from an order parameter part $F_Q({\bQ})$, an order
parameter-strain coupling $F_{Q\beps}(\bQ, \beps;\bX(P))$ and a 
purely elastic free energy term $F_{\eps}(\beps;\bX(P))$ (Toledano and Toledano, 1988). 
The order parameter part 
$F_Q({\bQ})$ is usually written as a polynomial in the variables ${\bQ}$ which is 
constructed using the theory of invariants of irreducible representations of the space groups
(Kovalev, 1993). 
The construction of the coupling term $F_{Q\beps}(\bQ, \beps;\bX(P))$ is more tricky. 
Measuring the strains with respect to the floating background reference system $\bX(P)$
and assuming linear-quadratic coupling of strain and order
parameter with coupling coefficients $d^{IJ}_{ij}(\bX(P))$, we write: 

\begin{eqnarray}
\label{Eq1}
F(\bQ,\widehat\beps;\bX(P))&=& V(P)\Phi(\bQ;\bX(P))
+V(P)\sum_{IJij}d^{IJ}_{ij}(\bX(P))Q_IQ_J\widehat\eps_{ij}
\nonumber
\\&&+ F_0(\widehat\beps;\bX(P))   
\end{eqnarray}

HPPT's are characterized by the fact that the total strain $\bm{\eta}$ is not small.
However, for phase transitions close to second order there is at least a nonzero
pressure interval starting at $P_c$ where the \emph{spontaneous} strain ${\widehat\beps}$ induced by the 
nonzero equilibrium order parameter $\bm{{\bar Q}}$ can still be treated as infinitesimal.  
In this approximation the pure elastic free energy part of the Landau potential with respect to $\bX(P)$
reads: 

\begin{eqnarray}
\label{Eq2}
F_0(\widehat\beps;\bX(P))\approx
V(P)\left(
\sum_{ij}\tau_{ij}\widehat\eps_{ij}
\frac{1}{2}\sum_{ijkl}C_{ijkl}(\bX(P))\widehat\eps_{ij}
\widehat\eps_{kl}
\right) 
\end{eqnarray}

where $C_{ijkl}(\bX(P))$ are the crystal's thermodynamic elastic constants at pressure $P$.
Inserting Eq.\ref{Eq2}  into Eq.\ref{Eq1}  and minimizing with respect to 
${\widehat\eps_{ij}}$ yields:

\begin{eqnarray}
\sum_{KL}d^{KL}_{ij}(\bX(P))\bar Q_K\bar Q_L+\sum_{kl}C_{ijkl}
(\bX(P))\widehat\eps_{kl}(\bar\bQ)\equiv0 
\label{Eq3}
\end{eqnarray}

Solving Eq.\ref{Eq3} for ${\widehat\eps_{ij}}(P)$ and inserting 
into the equilibrium
equation for the order parameter 

\begin{eqnarray}
0&\equiv&
\frac{\partial \Phi(\bar \bQ;\bX(P))}{\partial\bar Q_K}+
2\sum_{L}d^{KL}_{ij}(\bX(P))\bar Q_L\widehat\eps_{ij}(\bar\bQ)
\label{Eq4}
\end{eqnarray}

one obtains the renormalized order parameter potential written with 
respect to the 
deformed state $\bX(P)$:

\begin{eqnarray}
\lefteqn{\Phi_{R}(\bQ;\bX(P)):=\Phi(\bQ;\bX(P))-}\nonumber\\&&
-\frac{1}{2}\sum_{IJKL} Q_I Q_J Q_K  Q_L\left(
\sum_{ijkl}d^{IJ}_{ij}(\bX(P))C^{-1}_{ijkl}(P)d^{KL}_{kl}(\bX(P))\right)  
\label{Eq5}
\end{eqnarray}

In the spirit of Landau theory we now make the simple but crucial assumption that
the Landau and coupling coefficients of the potential depend on $P$ only in a ``geometrical'' way,
i.e.~through volume ratios and geometrical transformation rules between the reference states $\bX(P)$
and $\bX(0)$, such that the $P$ dependence of the equlibrium order pararameter is due to the presence of
the order parameter-strain coupling.
Transforming Eq.\ref{Eq5} to the undeformed reference 
state ${\bX(0)}$, the equilibrium order parameter 
${\bar \bQ}$ can then be calculated as the minimum of 

\begin{eqnarray}
\Phi_{R}(\bQ;\bX(0))=\Phi (\bQ;\bX(0))+\sum_{IJij}d_{ij}^{IJ}(\bX(0))e_{ij}
(P)Q_IQ_J- \nonumber
\\-\sum_{\stackrel{IJKL}{ijkl}}d_{ij}^{IJ}(\bX(0))\frac{T_{ijkl}(P)}{2}d_{kl}^{KL}(\bX(0))Q_IQ_JQ_KQ_L 
\label{Eq6}
\end{eqnarray}

where 

\numparts
\begin{eqnarray}
T_{rost}(\bX(P))&\equiv&
\frac{V(0)}{V(P)}
\sum_{ijkl}\al_{ir}\al_{jo}C^{-1}_{ijkl}(\bX(P))\al_{ks}
\al_{lt} \label{Eq7b}
\\
d_{mn}^{IJ}(\bX(P))&\equiv&\frac{V(0)}{V(P)}\sum_{ij}\al_{mi}d_{ij}^{IJ}(\bX(0))\al_{nj} 
\label{Eq7b}
\end{eqnarray}
\endnumparts

and the total (Lagrange) strain $\eta_{ij}$ turns out to be

\begin{eqnarray}
\eta_{ij}=e_{ij}
-\sum_{KL}\bar Q_K\bar Q_L\sum_{kl}d^{KL}_{kl}(\bX(0))T_{ijkl}(P)
\label{Eq8}
\end{eqnarray} 

Applying the above procedure and employing the usual assumptions concerning the
temperature behaviour of the Landau parameters, one is able to calculate 
pressure and temperature 
dependencies of thermodynamic quantities like specific heat, strains and 
elastic constants, soft mode behaviour, etc.~near HPPTs. 
A concrete example of the above theory at work will be given below.  
However, the above constructions obviously rely on the knowledge 
of the pressure dependence of the elastic constants $C_{ijkl}(\bX(P))\equiv C_{ijkl}(P)$ 
(or equivalently, the Birch coefficients 
$B_{ijkl}(P)=P(\delta_{ij}\delta_{kl}-\delta_{il}\delta_{jk}-\delta_{ik}\delta_{jl})+C_{ijkl}(P)$)
and the resulting background strains $e_{ij}(P)$. Note that these quantities
appear also in the renormalized free energy density Eq.\ref{Eq6}. 
To calculate it we proceed in the following way.

For crystal classes that develop no shear strains under hydrostatic pressure (cubic, tetragonal,
hexagonal, orthorhombic) the deformation tensor $\al_{ij}(P)=\al_i(P)\delta_{ij}$ is diagonal
and the axial compressibilities satisfy the equations

\begin{eqnarray}
-\frac{1}{\al_i(P)}\frac{d\al_i(P)}{dP}=\sum_{k}S_{ik}(P)=:\kappa_{i}(P) \qquad (i=1,2,3)  
\label{Eq9}
\end{eqnarray}

with the initial conditions $\al_i(0)=1$, where $S_{ij}(P):=B^{-1}_{ij}(P)$ denote the compliance tensor elements
and we switched to Voigt notation. 
Therefore the deformation tensor components $\alpha_i(P)$
as well as the finite strain tensor components $\eps_{ij}(P)$
or lattice parameters $a_i(P)$ can be easily calulated by integrating 
Eq.\ref{Eq9}, once the functions $S_{ik}(P),\ i,k=1,2,3$ are known.

In our recent work (Tr\"oster, Schranz and Miletich, 2002) we proposed the expansion

\begin{eqnarray}
S_{ij}(P) = \frac{\kappa(P)}{\kappa_0} \left( S_{ij}^0 + \sum_{n=1}^{\infty}
\kappa_{ij}^nP^n \right)  
\label{Eq10}
\end{eqnarray}

where $\kappa (P)$ is the bulk compressibility, $\kappa_0 = \kappa (P=0)$ and
$S_{ij}^0$ denotes the zero-pressure compliance.
The expansion coefficients $\kappa_{ij}^n$ obey the sum rule $\sum_{ij}\kappa_{ij}^n=0 \
\forall\ n\in\mathbb{N}$. Eq.\ref{Eq10},
which is exact at infinite $n$, yields an elegant parametrization of the pressure dependence of 
the compliance tensor, since it essentially factorizes out the main nonlinearity
of the compliance through the bulk compressibility $\kappa (P)$. 
The elastic anisotropy is taken into account by 
expanding the remainder in powers of $P$, as we anticipate that the very nature of the expansion
allows it to be truncated at low orders with good accuracy.
Indeed we have shown in a very recent work (Koppensteiner, Tr\"oster and 
Schranz, 2006) that even a low order truncation of our ansatz Eq.\ref{Eq10}
excellently fits the
experimental high pressure data of olivine, fluorite, garnet, magnesium oxide and stishovite. 
In fact, for all these examples a truncation at order $n=1$ turned out to be sufficient to produce accurate results,
the only exception being magnesium oxide, where the expansion Eq.\ref{Eq10}
had to be taken to $n=2$. This very encouraging result  
implies that our new thermodynamic theory indeed allows for a dramatic reduction 
of the number of fit parameters as compared to conventional nonlinear approaches. 
Table\ref{table1} shows the numbers of relevant (nonshear) elastic constants of second,
third and fourth order for various point group symmetries. Our approach
requires only $q_2-1$ additional unknown parameters $\kappa_{ij}^1$ (at order $n=1$), 
whereas a fit of an expansion of the elastic energy including nonlinear
elastic constants up to fourth order introduces $q_3+q_4$ additional 
fit parameters. For example, in the case of cubic I symmetry, our new
 approach requires $q_2-1=1$ 
fit parameters as compared to $q_2+q_3=7$ unknown parameters of a conventional fourth order
nonlinear theory. For lower symmetry the advantages of our approach are even more obvious.
For instance, in the case of orthorhombic symmetry  our new parametrization 
needs $5$ parameters, which should be 
compared to a total of $25$ unknowns for conventional nonlinear elasticity!

\begin{table}
\caption{\label{table1}
Numbers $q_2.q_3,q_4$ of nonshear elastic constants of second, third and 
fourth order (Roy and Dasgupta, 1988; Brendel, 1979).} 
\begin{tabular}{lllllllllllllll}
\br 
 & cubic I & cubic II & hexagonal I & hexagonal II & tetragonal I & 
tetragonal II & Orthorhombic\\
\mr
& $(432,{\bar 4}3m$ & $(23,m3)$ & $(622,6mm$ & $(6, {\bar 6}$ & $(422,4mm$
 & $(4,{\bar 4}$ & $(222,mm2$\\
& $m3m)$ &  &  ${\bar 6}m2, 6/mmm)$ & $6/m)$ & ${\bar 4}2m,4/mmm)$ & 
$4/m)$ & $mmm)$\\
\mr\\
$q_2$ & 2 & 2 & 4 & 4 & 4 & 4 & 6 \\
\mr\\
$q_3$ & 3 & 4 & 6 & 6 & 6 & 7 & 10 \\
\mr\\
$q_4$ & 4 & 5 & 13 & 13 & 9 & 9 & 15\\
\br
\end{tabular}
\end{table}

The combination of high pressure Landau theory and the use of the expansion 
(\ref{Eq10})  for
the background system $\bX(P)$ is illustrated by an
analysis of measurements of tetragonal
$\mathrm{BaCr(Si_4O_{10})}$ single crystals. 
Using x-ray diffraction, the
$P$-dependence of lattice parameters $a_1(P)=a_2(P)$, $a_3(P)$, 
and the unit cell
volumes $V(P)$ were measured very detailed at room
temperature in a diamond anvil cell (Tr\"oster, Schranz and Miletich, 2002). 

\begin{figure} 
\includegraphics[scale=0.5,angle=-90]{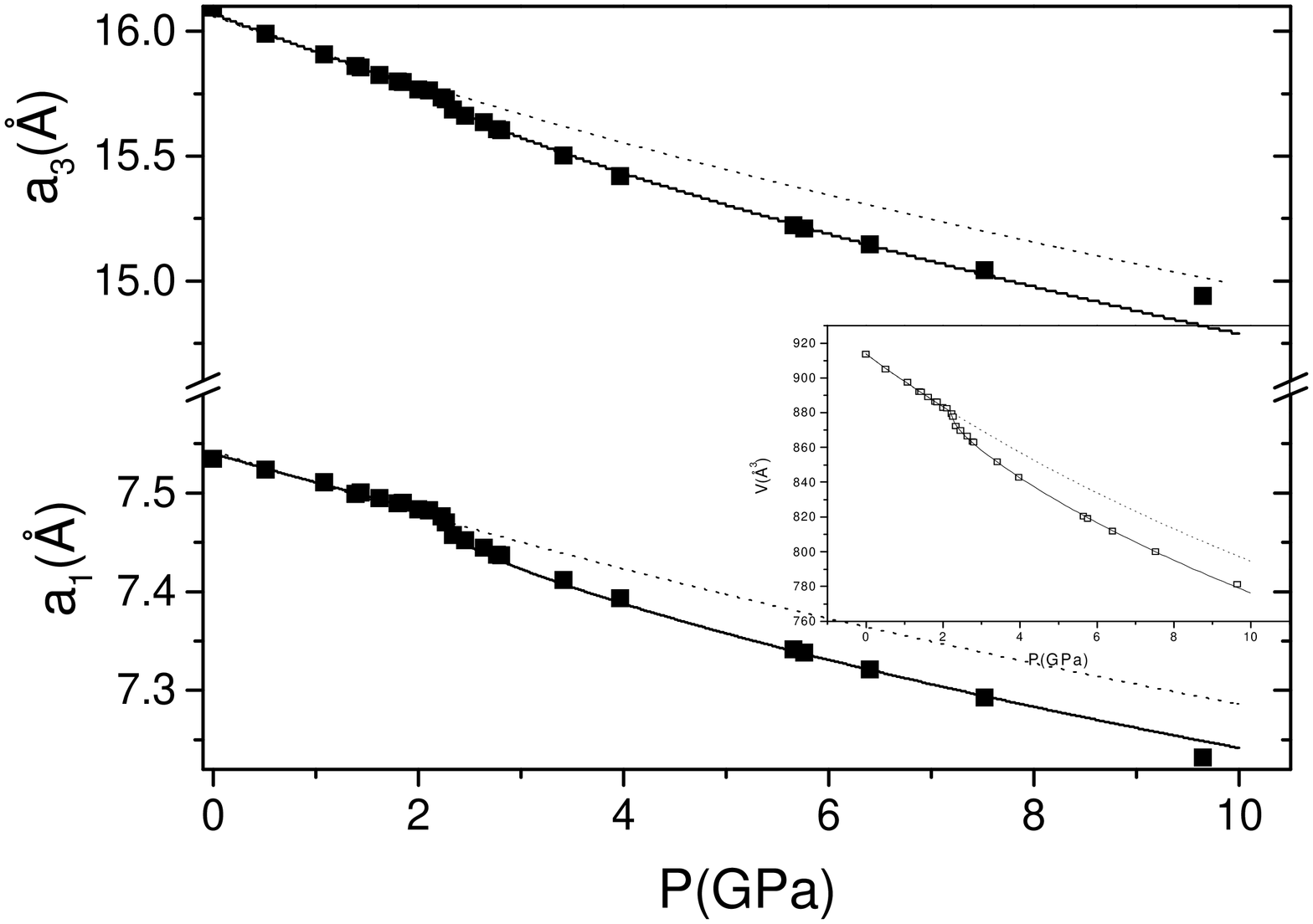}
\caption{Pressure dependence of lattice parameters and unit cell volume of 
$\mathrm{BaCr(Si_4O_{10})}$. Points are measured data, lines are fits 
using the present theory.}
\label{Fig2}
\end{figure}

\begin{table}
\caption{\label{table2}
Fit parameters for the high pressure Landau theory of 
$BaCr(Si_4O_{10})$.}
\begin{indented}  
\item[]\begin{tabular}{ll}
\br
&  $BaCr(Si_4O_{10})$     \\
\br\\
$A$                   & 0.45 GPa               \\
$B$                   & -0.2 GPa             \\
$C$                   & 20 GPa              \\
$a_1(0)=a_2(0)$       & 7.535 \AA              \\
$a_3(0)$              & 16.09 \AA              \\
$\kappa_1^0$         & 0.0035 GPa$^{-1}$      \\
$\kappa_3^0$               & 0.001 GPa$^{-1}$      \\
$S_{11}^0$            & 0.0035 GPa$^{-1}$       \\
$S_{3}^0$             & 0.0089 GPa$^{-1}$       \\
$K_0'$                & 4.1                  \\
$\kappa_{11}^1$   & 1.9$\times10^{-4}$ GPa$^{-2}$  \\
$\kappa_{12}^1$   & -1.7$\times10^{-4}$ GPa$^{-2}$    \\
$\kappa_{13}^1$   & 2.3$\times10^{-4}$ GPa$^{-2}$      \\
$d_1$                 & 13.75 GPa$^{-1}$      \\
$d_3$                 & -0.33 GPa$^{-1}$      \\
\br
\end{tabular}
\end{indented}
\end{table}

One finds a tetragonal-tetragonal HPPT at approximately 
 $2.24$ GPa,  characterized by a
discontinuitiy in $a_1(P)$, $a_3(P)$ and $V(P)$.  
Consistent with the observed pressure hysteresis  
behavior, the transition can be classified as being of (weakly) first order.
The order parameter part is  constructed in  the following 
standard way (Kovalev, 1993): The symmetry reduction $\mathrm{P4/ncc}$
to $\mathrm{P42_12}$ is driven by the onedimensional irreducible representation
$\tau_2$ at the wavevector $\bk=\bnull$, yielding a   
one component order parameter $Q$, which is zero in the paraphase 
$(P<P_c)$ and nonzero in the distorted phase $(P>P_c)$.
For $\Phi(Q,{\bX})$ we assume

\begin{eqnarray} 
\Phi(Q,{\bX})=
\frac{A}{2}Q^2+\frac{B}{4}Q^4+\frac{C}{6}Q^6  
\label{Eq11}
\end{eqnarray}

where $A, C>0$.
The tetragonal symmetry also dictates $d_1({\bX)}=d_2({\bX})\neq d_3({\bX})$.

Let $K_0:=\kappa^{-1}(0)$ denote the isothermal bulk modulus at $P=0$.
The Murnaghan equation of state (MEOS) (Anderson, 1995) 

\begin{eqnarray}
v(P):= \frac{V(P)}{V(0)} = \left(1+K'_0 P/K_0\right)^{-1/{K_0'}}
\label{Eq12}
\end{eqnarray}

which is based on the simple ansatz $\kappa^{-1}(P)=:K(P)=K_0+K_0' P$, 
is frequently used to describe $(P,V)$-data and is known to usually reproduce the values of
$K(P)$ correctly up to volume changes somewhat larger than $v(P)>0.9$ while being algebraically
much simpler than other approaches such as the 
"Vinet" or the "Birch-Murnaghan" EsOS (Anderson, 1995; Angel, 2000b)  
used for higher compression ranges. 
Fig.\ref{Fig2} shows corresponding fits of unit cell
volume and axes of $\mathrm{BaCr(Si_4O_{10})}$ using the parameter 
values of Table\ref{table2}. Note, however, that our theory can in principle be used in combination with any 
of these EsOS, or even a function $V(P)$ derived from an experimental measurement or a 
computer simulation. With these values our model confines possible pressure ranges for hysteresis effects to
2.2 GPa--2.4 GPa. One also calculates that the geometrical error introduced 
in assuming the spontaneous strain ${\widehat\beps}$ to be infinitesimal is 
smaller than 0.9\%, yielding  an error $<0.1\%$ in the total strain 
${\bf \eta}$.

\begin{figure}[h] 
\includegraphics[scale=0.5,angle=-90]{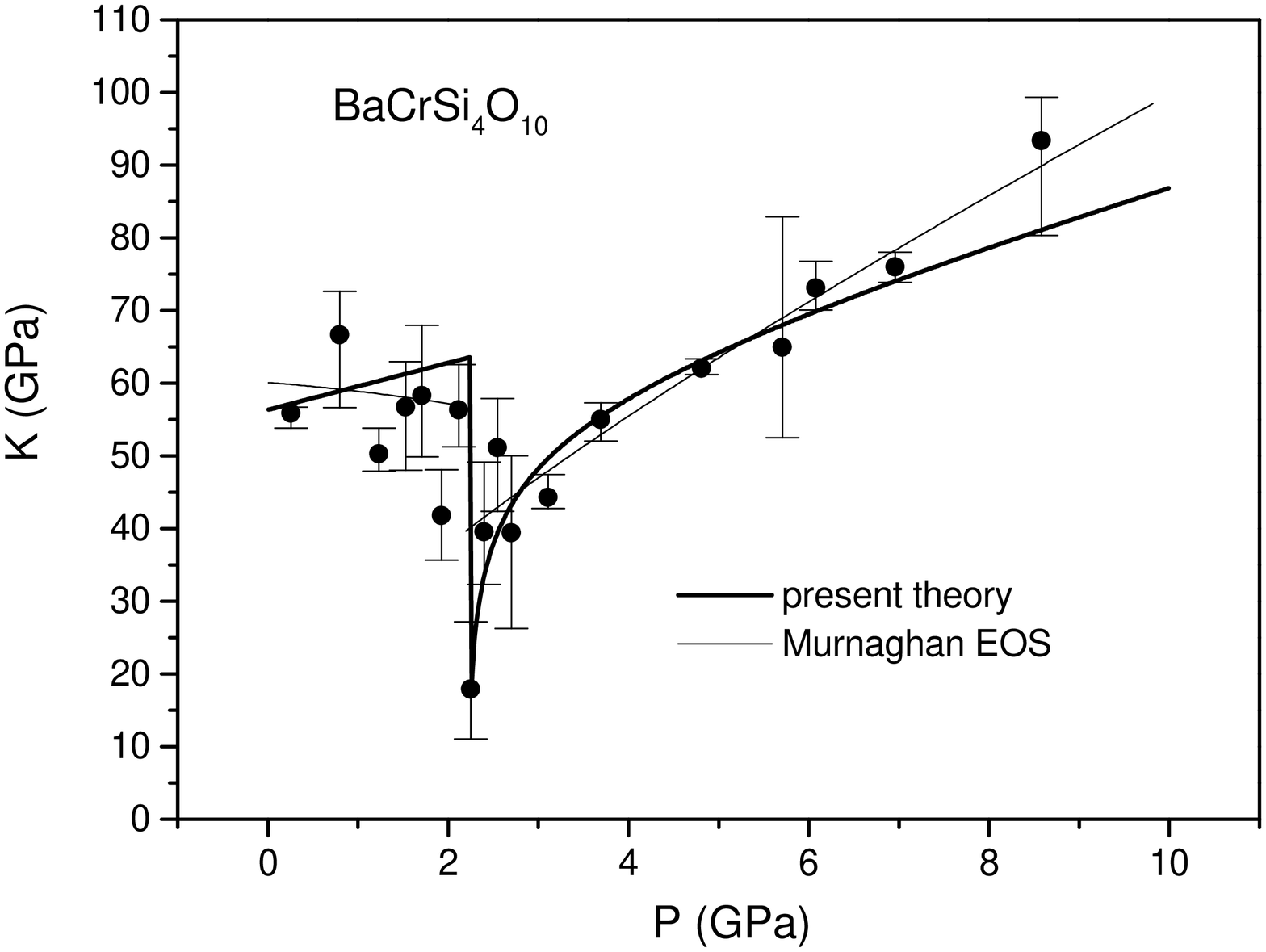}
\caption{Pressure dependence of bulk modulus $K(P)$ calculated from the 
experimental data of $V(P)$ (points). The lines show calculations based on 
piecewise EOS fitting (thin line) and the present high pressure Landau theory
(thick line).} 
\label{Fig3}
\end{figure}

\begin{figure}[h]
\includegraphics[scale=0.5,angle=-90]{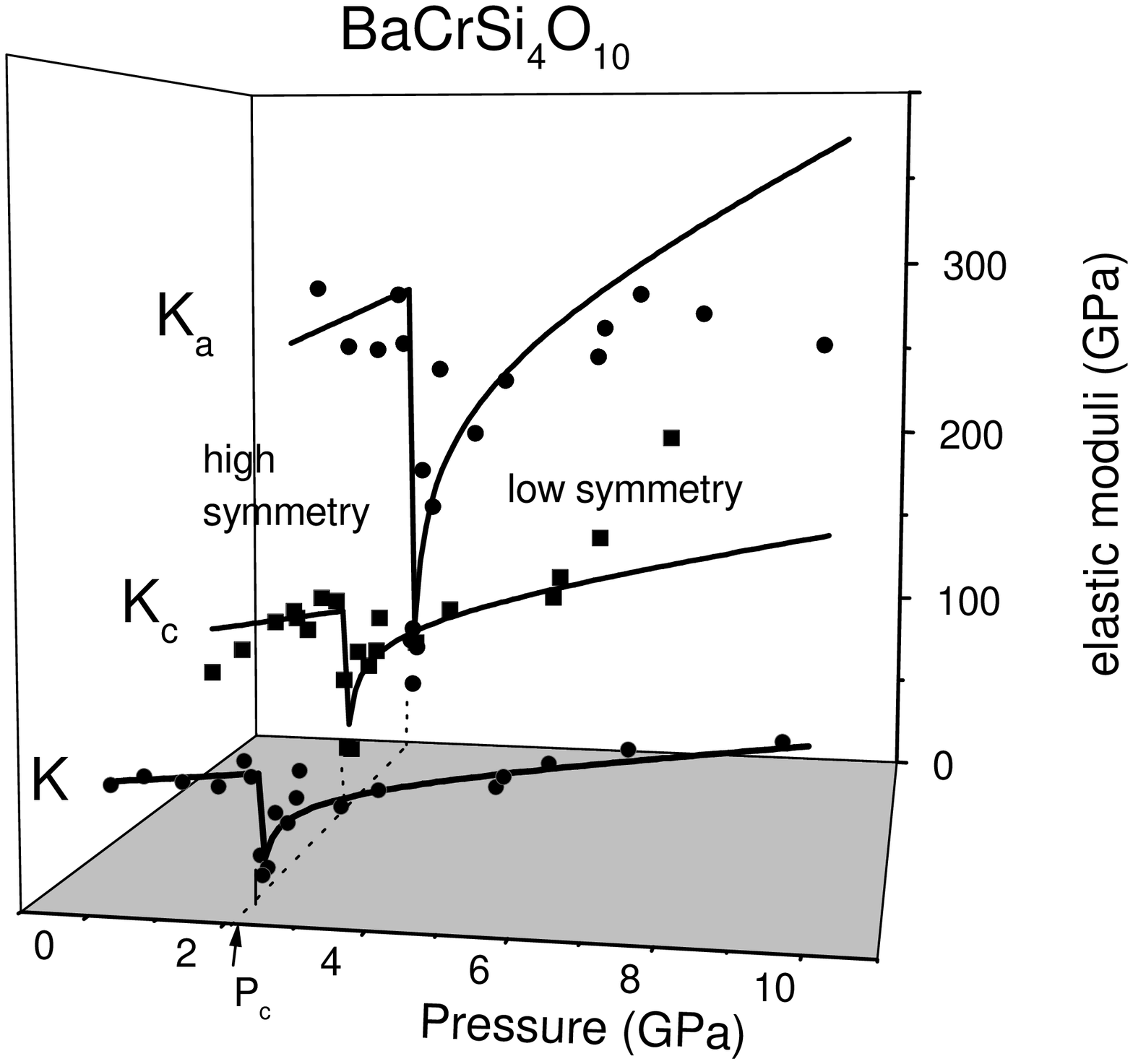}
\caption{Pressure dependence of axial incompressibilites $K_a, K_c$
calculated from the
lattice parameters $a_1, a_3$ (points) and the bulk modulus $K$
compared with the results of the present 
Landau theory (lines).}
\label{Fig4}
\end{figure}

Figure \ref{Fig3} shows the pressure dependence of the bulk modulus of $BaCr(Si_4O_{10})$ calculated from 
the experimental data of $V(P)$ as compared to the predictions of the different approaches discussed above. 
It is clearly evident that the conventional fitting procedure, which is based on 
a piecewise fitting of EOS in the high and low pressure phase, respectively
, underestimates the elastic anomaly by at least a 
factor of two, whereas the present high pressure adapted Landau theory 
reproduces the elastic anomaly very well. 
The same behaviour is found for the axial imcompressibilities (Fig.\ref{Fig4}). 
It is worth noting that a very similar softening was also observed in other examples of
pressure-induced phase transitions, notably 
for the bulk modulus of solid $C_{60}$ near its fcc-sc transition (Pintschovius, 1999)
and for the longitudinal acoustic modes near the cubic-tetragonal transition of $BaTiO_3$
(Ishidate, 1989). 

\section{Summary}

Summarizing, our theory allows to compute a wealth of experimental observables
(e.g. lattice parameters, elastic constants, specific heat, soft modes, etc.)
from a quite transparent thermodynamic model based on
coupling finite strain elasticity to Landau theory. There are several advantages 
of this new approach to HPPTs:
At first there is a drastic reduction of fit parameters
as compared to an expansion in terms of nonlinear elastic constants
(Table\ref{table1}). 
Moreover the present approach allows for a direct connection 
to EOS fitting procedures. Compared 
to the method of piecwise EOS fitting the main advantage of our method is that it
allows for a unified consistent description of high and low symmetry phases with a single
set of parameters. \\
At present, our theory is capable of dealing with group-subgroup transitions involving 
symmetries of cubic, tetragonal, orthorhombic or hexagonal type. However, it is possible to
generalize our approach to arbitrary symmetries once the pressure dependence of the shear
elastic constants (in addition to that of the longitudinal ones) is known. Further work in this direction
is in progress. 

\ack 
Support by the Austrian FWF (P19284-N20) and the {\it University of Vienna} 
(Initiativkolleg IK 1022-N) is gratefully acknowledged.  

\section*{References}
\begin{harvard}

\item[Anderson, O.L. (1995).] \textit{Equations of State of Solids
    for Geophysics and Ceramic Sciences}, Oxford University Press, Oxford.
\item[Angel, R.J. (2000a).] \textit{High-Pressure Structural Phase
    Transitions}, in  "Transformation Processes in Minerals", Reviews
    in Mineralogy and Geochemistry, 
    Vol.~\textbf{39}, p.~85, ed. S.A.T.~Redfern and M.A.~Carpenter.
\item[Angel, R.J. (2000b).] \textit{Equations of state}, In R.M.~Hazen (ed),
\textit{High-Temperature-High-Pressure Crystal Chemistry}, Reviews in
    Mineralogy \textbf{40}. 
\item[Bergman, D.J. and Halperin, B.I., (1976).]
%Critical behavior of an Ising model on a cubic compressible lattice.
Phys. Rev. B {\bf 13}, 2145-75.
\item[Bratkovky, A.M., Salje, E.K.H., Marais, S.C., and Heine, V. (1995).]
%Strain coupling as the dominant interaction in structural phase transitions.
Phase Transitions {\bf 55}, 79-126.
\item[Brendel, R. (1979).]
% Independent Fourth-Order Elastic Coefficients for All 
Crystal Classes. Acta Cryst. A{\bf 35}, 523-533. 
\item[Carpenter, M.A., Salje, E.K.H. and Graeme-Barber, A. (1998).]
%Spontaneous strain as a determinant of thermodynamic properties for 
%phase transitions in minerals. 
Eur. J. Mineral., {\bf 10}, 621-691.
\item[Chesnut, G.N. and Vohra, Y.K. (2000).]
%Phase transformations and equation of state of praesodymium metal to 103 GPa. 
Phys. Rev. B \textbf{62}, 2965-68.
\item[Dove, M.T. (1997).] 
%Theory of displacive phase transitions in minerals. 
American Mineralogist, {\bf 82}, 213-244.
\item[Giddy, A.P., Dove, M.T. and Heine, V. (1989).]
%What do Landau free energies really look like for structural phase 
%transitions?
J.Phys.: Condens. Matter, {\bf 1}, 8327-8335. 
\item[Ishidate, T., and Sasaki, S. (1989).]
%Elastic anomaly and phase transition of $BaTiO_3$, 
Phys.~Rev.~Lett.~{\bf 62}, 67-70.
\item[Koppensteiner, J., Tr\"oster, A. and Schranz, W. (2006).] 
%Efficient parametrization of high-pressure elasticity. 
Phys. Rev. B {\bf 74}, 014111. 
\item[Kovalev, O.V. (1993).] \textit{Representation of the Crystallogra\-phic
Space Groups}, edited by H.T.~Stokes and 
D.M.~Hatch, Gordon and Breach Science Publishers.
\item[Kr\"uger, T.  and Holzapfel, W.B. (1992).]
%Structural phase transitions and equations of state for selenium 
%under pressures to 129 GPa. 
Phys. Rev. Lett. \textbf{69}, 305-307.
\item[Meyer, H.W., Carpenter, M.A., Graeme-Barber, A., Sondergeld, P. and 
Schranz, W. (2000).] Eur. J. Mineral. {\bf 12}, 1139. 
\item[Pérez-Mato, J.M., Ivantchev, S., García, A. and Etxebarria, I. (2000).] 
%Displacive vs. order-disorder in structural phase transitions. 
Ferroelectrics, {\bf 236}, 93-103.
\item[Pintschovius, L., Blaschko, O., Krexner, G. and Pyka, N. (1999).]
%Bulk modulus of $C_{60}$ studied by single-crystal neutron diffraction. 
Phys. Rev. B \textbf{59}, 11020-26.
\item[Roy, D. and Dasgupta. (1988).]
 \textit{Lattice Theory of Elastic Constants}, 
edited by S. Senegupta (Transtech Publications Ltd., Switzerland). 
\item[Rubtsov, A.N., Hlinka, J. and Janssen, T. (2000).]
%Crossover between a displacive and an order-disorder phase transition. 
Phys. Rev. E, {\bf 61}, 126-131. 
\item[Salje E.~K.~H. (1990).] \textit{Phase Transitions in
    Ferroelastic and Coelastic Crystals}, Cambridge University Press, 
pp.1-229.
\item[Salje, E.K.H. (1992).]
%Application of Landau theory for the analysis of phase 
%transitions in minerals. 
Physics Reports, {\bf 215}, 49-99.
\item[Schulte, O. and Holzapfel, W.B. (1995).]
%Equation-of-state behavior for different phases of lead under 
%strong compression. 
Phys. Rev. B, \textbf{52}, 12636-39.
\item[Sollich, P., Heine, V. and Dove, M.T. (1994).]
%The Ginzburg interval in soft-
%mode phase transitions: consequences of the rigid unit mode picture. 
J.Phys.: Condens. Matter, {\bf 6}, 3171-3196.
\item[Sondergeld, P., Schranz, W., Tr\"oster, A., Carpenter, M.A., 
Libowitzky, E. and Kityk, A.V. (2000)]. Phys. Rev. B{\bf 62}, 6143. 
\item[Tr\"oster, A., Schranz, W. and  Miletich, R. (2002).]
%How to Couple Landau Theory to an Equation of State. 
Phys. Rev. Lett., {\bf 88}, 055503,1-4.
\item[Tr\"oster, A., Dellago, C. and Schranz, W. (2005).]
%Free energies of the $\phi^4$-model from Wang-Landau simulations. 
Phys. Rev. B, {\bf 72}, 094103,1-11.
\item[Wallace, D.C.(1972).]  \textit{Thermodynamics of Crystals},
John Wiley \& Sons, Inc., U.S.A.

\end{harvard}
\end{document}